# A semi-analytical model for the scale-dependent friction of nanosized asperity

Jian Wang, Weike Yuan, Jianjun Bian and Gangfeng Wang[*]

*Department of Engineering Mechanics, SVL, Xi'an Jiaotong University, Xi'an 710049, P. R. China.*

[*]E-mail: wanggf@mail.xjtu.edu.cn

## Abstract

The friction of a nanosized sphere in commensurate contact with a flat substrate is investigated by performing molecular dynamics simulations. Particular focus is on the distribution of shear stress within the contact region. It is noticed that within the slip zone, the local friction coefficient defined by the ratio of shear stress to normal pressure declines monotonically as the distance to contact center increases. With the lateral force increasing, the slip zone expands inwards from the contact edge. At the same time, the local friction coefficient at the contact edge decreases continuously, while at the dividing between the slip and stick zones keeps nearly invariant. These characteristics are distinctly different from the prediction of the conventional Cattaneo-Mindlin model assuming a constant local friction coefficient within the slip zone. An analytical model is advanced in view of such new features and generalized based on numerous atomic simulations. This model not only accurately characterizes the interfacial shear stress, but also explains the size-dependence of static friction of single nanosized asperity.

# 1. Introduction

For the macroscopic dry friction between two sliding solids, the classical Amontons-Coulomb's law is still the most important understanding to date. This phenomenological friction theory states that the lateral friction force is always proportional to the normal load by a constant friction coefficient, which is independent of the contact area, surface roughness and sliding velocity [1]. Although such linear characteristic was originally summarized from a global aspect, it has been widely employed to correlate the local shear stress with the corresponding contact pressure at local slipping spots [2].

In the framework of the well-known Cattaneo-Mindlin (CM) model, relative motion occurs at the point of contacting interface where the shear stress exceeds the local strength equaling the normal pressure times a constant local coefficient of friction [2],[3]. This classical model predicts a continuous transition from partial slip contact to overall sliding. When the entire contact region comes into slipping stage, the lateral friction force acting on the sliding body reaches maximum. Moreover, the maximum friction force equals the normal load multiplied with a global coefficient of friction, which is consistent with the local one. On the other hand, for the adhesive contact problem, the assumption of pressure independent shear stress is commonly adopted in the slip zone [7][9]. In this case, the friction force is equal to the real contact area times a constant shear strength. These analytical models have been applied to characterize the friction process of some single asperity experiments [10][13].

When the contact scale comes to nanometers, the aforementioned continuum analyses always break down because some unconventional features have not been involved, e.g., the repeatedly reported scale-dependent frictional behavior [14][20]. To study the scale effects in single asperity friction, Hurtado and Kim [16] proposed a dislocation gliding model to calculate the friction force. It was shown that when the contact area is smaller than a critical value, the interfacial slip is concurrent with a constant friction stress. However, for larger contact size, overall sliding advances when a dislocation loop sweeps through the contact region, and the friction stress would decrease with the contact area increasing. By using the Peierls-Nabarro model of interface dislocation field, Gao presented the size dependence of single asperity friction that evolves from homogeneous slip mode to dislocation-slip-like inhomogeneous slip mode [18]. Recently, for the non-adhesive commensurate contact between a rigid sphere and a flat elastic substrate, Sharp *et al.* calculated the friction force as a function of contact scale by using a Green's function molecular dynamics method [19]. Their results revealed that the static friction coefficient is dependent on the contact area, which is constant at small contact size and fits to a $-2/3$ power law relation at larger contact size.

In this work, the frictional behavior of single spherical asperity in commensurate contact with a flat substrate is simulated by using complete molecular dynamics (MD) with purely repulsive tip-substrate interaction. Scale-dependent frictional behavior as displayed in the previous study [19] is confirmed. More importantly, details of the distribution of shear stress within the contact region are examined carefully. It is found that in the slip zone, the local friction coefficient defined by the ratio of shear stress to normal pressure essentially decreases with the distance to contact center increasing, quite different from the existing frictional models using constant local friction coefficient [2][3]. Such new characteristic of the local shear stress is generalized based on atomic simulation results, which gives rise to an accurate analytical model for the scale-dependent friction of single asperity at nanoscale.

## 2. Simulation Method

Fig. 1 depicts the model used in present study. The substrate is made of single crystal copper with the dimension of $22\times22\times11$ nm$^3$. Substrates with larger size are also examined, and no significant difference is observed in the simulation results. To lower the computational burden, this sized substrate is adopted for all the following simulations. The lattice orientations along the *x*-, *y*- and *z*-directions are [110], [001] and [1$\bar{1}$0], respectively. A rigid spherical tip with radius $R$ is created by bending a thin atomic layer with the same lattice orientations as the substrate. Thus, the contacting surfaces are specified as [001], and commensurate contact can be achieved between the tip and substrate. Periodic boundary conditions are applied along both $x$- and $z$-directions, and the 0.5 nm-height atom layer at the bottom of the substrate is fixed in perfect lattice.

The embedded atom method (EAM) [21] is utilized to describe the atomic interactions in the substrate. Based on EAM, for an atomic system containing $N$ atoms, the energy of the $i$-th atom is given by

$$E_i = F(\bar{\rho}_i) + \frac{1}{2}\sum_{i \neq j}^{N} \phi(r_{ij}), \tag{1}$$

where $F(\bar{\rho}_i)$ is the embedded energy related to the electron density $\bar{\rho}_i$ at the position of the $i$-th atom, $\phi(r_{ij})$ is the pair potential interaction, and $r_{ij}$ represents the distance between atom pair $i$ and $j$. A potential of copper parameterized by Mishin *et al.* [22] is adopted here, which has been widely used to investigate mechanical properties and deformation mechanisms of various nanostructures [23].

To implement the non-adhesive contact, the atomic interaction between two opposing contact surfaces is described through a repulsive Lennard-Jones potential,

$$E = 4\varepsilon\left[\left(\frac{\sigma}{r_{ij}}\right)^{12} - \left(\frac{\sigma}{r_{ij}}\right)^{6}\right] + \varepsilon, \ r_{ij} < 2^{1/6}\sigma, \tag{2}$$

where $\varepsilon$ governs the interaction intensity, and $\sigma$ is the equilibrium spacing. The values are calculated based on the cohesive energy and the lattice structure of single crystal copper, which yields $\varepsilon = 0.423$ eV and $\sigma = 2.325$ Å.

The large-scale atomic/molecular massively parallel simulation package, LAMMPS, is adopted to conduct the simulations [24], and the time evolution of the atomic system is implemented within the fra mework of canonical ensemble (NVT). After the construction of the atomic model, energy relaxation is first performed based on a conjugate gradient algorithm to get an energy-favorable initial state. To exclude the thermal effects, temperature in the simulations is set as 0.001 K. After the static energy relaxation, the atomic system is dynamically equilibrated at 0.001 K for 50 ps. Then, a loading scheme similar as in [25] is implemented to simulate the contact and friction process. A small vertical displacement $u_y$ is gradually exerted on the tip resulting in a circular contact region with radius $a$. The corresponding normal load $P$ is computed by summing up the normal pressure within the contact area. In the following friction stage, the tip is further pulled along the $x$-direction with a constant speed of 0.01 Å/ps.

To analyze the friction process at atomic scale, the force state of each atom in the substrate is traced. For the tangential force exerted on every single atom, we only consider the component along the sliding direction ($x$-direction). It should be pointed out that even only normal load is applied, there is some local shear stress on the contact surface. This initial atomic force state before the tip being pulled laterally is chosen as a reference. When a specified lateral displacement $u_x$ is applied on the tip, the incremental portion of the tangential force of each atom is defined as the atomic friction force. Summing up the atomic friction force over all contacting atoms yields the total lateral force on the tip $F$. Accordingly, the local shear stress at a specific atom site is obtained by dividing the atomic friction force with the average contact area of a single atom. At each site, the ratio of the local shear stress to the local normal pressure defines the local friction coefficient. In addition, the relative tangential displacements between the contacting surfaces of the tip and substrate are examined. The contact area is divided into a central stick zone and a surrounding slip zone according to whether the relative displacement is zero or not.

## 3. Results and discussions

Fig. 2 displays the variation of the lateral force on the tip with radius $R = 200$ nm, normalized by the corresponding normal load, as a function of lateral displacement. For a given normal load, the lateral force first increases monotonically as the tip moves along the $x$-direction. By examining the atomic displacements on the contact surface, it is found that in the initial stage the slip zone broadens continuously with the lateral displacement increasing, while the inner stick zone shrinks inwards. After reaching a peak, the lateral force drops gradually till to a minimum with further tip lateral displacement. The peak lateral force is defined as the static friction force of the single asperity. Dividing the static friction force by the relevant normal load derives the static friction coefficient. Notably, it is found that the static friction coefficient, i.e., the peaks of the curves in Fig. 2, declines with the increasing of normal load. In addition, the appearance of the peak lateral force is also postponed. Such load- or scale-dependency of static friction was analyzed based on the assumption of a constant local friction coefficient in the slip zone [19].

Fig. 3a displays the distribution of local shear stress on the contact surface under normal load $P = 1.2$ keV/Å. A rotationally symmetric feature is observed. To get more quantitative analysis, the local shear stress along five directions ($\theta = 0°, 45°, 90°, 135°, 180°$) is plotted in Fig. 3b. The well reserved rotation symmetry could be used as a simplification in the following discussion. It is seen that in the central stick zone, the shear stress steadily increases with the distance to contact center increasing and reaches to a peak at the boundary between the stick and slip zones. In the slip zone, the stress declines gradually with the distance away from the center.

For comparison, the shear stress based on CM model under the same normal and lateral force is also plotted in Fig. 3b, which has the following expression,

$$\tau(r) = \begin{cases} \mu_m p_0 \left(1 - \frac{r^2}{a^2}\right)^{1/2} - \mu_m p_0 \frac{c}{a} \left(1 - \frac{r^2}{c^2}\right)^{1/2}, & r \leq c \\ \mu_m p_0 \left(1 - \frac{r^2}{a^2}\right)^{1/2}, & c < r \leq a \end{cases}, \quad (3)$$

where $r$ is the distance to the contact center, $\mu_m$ is a constant local friction coefficient and here takes the value of 0.7 as in [19], $p_0$ is the normal pressure in the center of the contact, $a$ is the contact radius, and $c$ is the radius of the stick zone.

In the CM model, the normal pressure in the contact region is assumed to obey the Hertzian theory. In MD simulations, the normal pressure also agrees well with the distribution of Hertzian pressure. Thus, the contact radius is determined by $a = (3PR/4E^*)^{1/3}$ when the normal load $P$, tip radius $R$ and indentation modulus $E^*$ are specified, and $p_0 = 3P/(2\pi a^2)$. Moreover, the radius ratio of the stick zone and contact region is given by $c/a = [1-F/(\mu_m P)]^{1/3}$ with $F$ being the lateral force. It is shown in Fig. 3b that CM model is unable to accurately describe the distribution of local shear stress. Under the same normal and lateral loading, CM model overestimates the size of stick zone. More importantly, the drop tendency of shear stress of CM model in the slip zone is remarkably different from that of MD simulations.

To advance the characterization of local shear stress, we further examine the local friction coefficient, i.e. the ratio of local shear stress to local normal pressure within the contact region. For a given normal load $P = 1.2$ keV/Å, we plot the local friction coefficient along the path $\theta = 0°$ in Fig. 4. It is shown that in the slip zone, the local friction coefficient declines with the increasing of the distance to contact center, rather than being constant. At the dividing boundary between stick and slip zones appears the maximum local friction coefficient, which approximately equals to the coefficient of static friction for a single atom [19]. As the lateral displacement increases, the maximum value of local friction coefficient within the slip zone keeps nearly unchanged, while the local friction coefficient at the contact periphery decreases gradually. Once the local friction coefficient at the contact edge reduces to zero, a dislocation will nuclear here and sweep across the whole contact surface rapidly, inducing a burst overall sliding.

Based on above characteristics, we find the local friction coefficient in the slip zone can be well described by,

$$\mu_l(r) = \frac{\tau(r)}{p(r)} = \mu_m + \beta\mu_m \left[\left(1 - \frac{r^2}{a^2}\right)^{\frac{3}{2}} - \left(1 - \frac{c^2}{a^2}\right)^{\frac{3}{2}}\right], \quad c < r \leq a, \qquad (4)$$

where $p(r)$ is the local normal pressure following the Hertzian prediction, $\mu_m$ is the maximum of local frictional coefficient and takes 0.7, and $\beta$ is a dimensionless parameter related to the normal load, the tip radius and the material properties. From Eq. (4), the local friction coefficient at the edge of contact region is derived as,

$$1 - \frac{\mu_l(a)}{\mu_m} = \beta\left(1 - \frac{c^2}{a^2}\right)^{3/2}. \qquad (5)$$

By implementing a series of atomic simulations for the indenter with radius $R = 200$ nm, the local friction coefficients at the contact edge under different normal load are captured. As shown in Fig. 5, good linear relationship between $1-\mu_l(a)/\mu_m$ and $(1-c^2/a^2)^{3/2}$ is observed, which means $\beta$ is independent of the lateral force. Next, we will attempt to derive the explicit expression of the parameter $\beta$, which is the key of accurate characterization of shear stress.

In existing study, it has been shown that the scale-/load-dependent frictional behavior can be uniquely determined by a dimensionless parameter defined by the ratio of contact radius to the minimum core width of the dislocation slipping across the contact interface [19]. Using the normal contact pressure predicted by Hertzian theory, the dimensionless parameter is expressed by

$$\frac{a}{b} = \left(\frac{9E^*P^2}{2R}\right)^{\frac{1}{3}} \frac{\mu_m}{\pi G d}, \tag{6}$$

where $b$ is the core width of interface dislocation, $G$ is shear modulus, and $d$ is the minimum lattice period along the sliding direction. In Fig. 6, the value of $\beta$ is plotted against the characteristic ratio $a/b$. The simulation results of indenters with radius $R = $ 100 nm and 150 nm are added. It is found that $\beta$ is approximately proportional to the ratio $a/b$. Direct linear fitting gives

$$\beta = 0.6 \cdot a/b = 0.6 \cdot \left(\frac{9E^*P^2}{2R}\right)^{\frac{1}{3}} \frac{\mu_m}{\pi G d}. \tag{7}$$

Now, with Eqs. (4) and (7), the local shear stress in the slip zone can be well characterized in a general and explicit manner. In the stick zone, we employ the expression in the same form as CM model to describe the distribution of shear stress. As a result, the shear stress in the whole contact area is expressed as

$$\tau(r) = \begin{cases} \mu_m p_0 \left(1 - \frac{r^2}{a^2}\right)^{1/2} - \mu_m p_0 \frac{c}{a}\left(1 - \frac{r^2}{c^2}\right)^{1/2}, & r \leq c \\ \mu_l(r) p_0 \left(1 - \frac{r^2}{a^2}\right)^{1/2}, & c < r \leq a \end{cases}, \tag{8}$$

where $\mu_l(r)$ is given by Eqs. (4) and (7).

Compared with our MD simulation results shown in Fig. 3, this modified shear stress distribution shows great accuracy in both stick zone and slip zone. When the tip radius is much larger or the normal force is much smaller, $\beta$ is negligible and the present model reduces to CM model. However, when the tip radius is relatively small and thus $\beta$ is appreciable, the present model will give a noticeable correction to the shear stress.

The total lateral force acting on the tip can be obtained by integrating the local shear stress over the entire contact region,

$$F = \mu_m P \left[1 - \frac{c^3}{a^3} - \frac{\beta}{2}\left(1 - \frac{c^2}{a^2}\right)^3\right], \tag{9}$$

where the third term in the bracket is the correction arising from the scale effects. Fig. 7 plots the normalized lateral force $F/(\mu_m P)$ against the normalized width of the slip zone $1-c/a$. All of our MD simulation results are in good agreement with the prediction of the scale-dependent model given by Eq. (9).

According to Eq. (9), the radius of stick zone associated with the maximum lateral force should be determined by $\partial(F/\mu_m P)/\partial(c/a) = 0$, which yields

$$\beta\left(1 - \frac{c^2}{a^2}\right)^2 - \frac{c}{a} = 0. \tag{10}$$

On the other hand, we must bear in mind that when the local friction coefficient at the contact edge decreases to zero, the overall sliding would start. At this critical point, the radius of the stick zone satisfies

$$\frac{c}{a} = \sqrt{1 - \beta^{-2/3}}. \tag{11}$$

To predict the maximum static friction force $F_{static}$, these two critical conditions Eq. (10) and (11) are compared in Fig. 8. There are two regimes for the determination of maximum static friction force. In regime I where $\beta < 2.83$, as the lateral displacement increases, the width of slip zone $1-c/a$ first reaches the red dash line given by Eq. (10). Thus, the maximum static friction force should be the maximum of Eq. (9). In contrast, in regime II where $\beta > 2.83$, the width of slip zone $1-c/a$ would stop at the blue dot-dash line given by Eq. (11). At this moment, the overall contacting sliding would initiate. The lateral force on the tip will not continue to increase any more. Thus, the maximum static friction force should be just the value of Eq. (9) with the radius of stick zone given by Eq. (11).

From Fig. 8, we can also recognize two different sliding modes. For $\beta < 1$, the local friction coefficient at contact edge is always positive since Eq. (11) will never be reached. Only when the stick zone shrinks to zero, does the overall sliding happens. On the other hand, for $\beta > 1$, overall sliding takes place once the local coefficient at contact age decreases to zero.

For $\beta > 2.83$ (i.e., in the regime II), based on the critical radius of stick zone from Eq. (11), the coefficient of global static friction is derived as

$$\mu/\mu_m = F_{static}/P = 1 - \left(1 - \beta^{-2/3}\right)^{3/2} - \frac{1}{2\beta}, \tag{12}$$

which is dependent only on the scale parameter $\beta$. By eliminating the high order terms, it gives

$$\mu/\mu_m \approx \frac{3}{2}\beta^{-2/3}. \tag{13}$$

In Fig. 9, the static friction coefficients of single asperities with different radius ($R = 100, 150, 200$ nm) under different normal load are displayed. Note that the scale of our complete atomic simulations is limited by the computation capacity. For further testifying our model, we also present the results of Sharp *et al*. [19]. It is found that the prediction of our scale-dependent model agrees well with the simulations results. For $\beta < 0.1$, the static friction coefficient $\mu$ is nearly constant and equals to the maximum local friction coefficient. For $\beta > 10$, $\mu$ varies as the $-2/3$ power of $\beta$ as described by Eq. (13), which is consistent with the scaling results in [19]. In addition, the transition between these two regimes is successfully characterized. It should be pointed out that Sharp *et al*. [19] also calculated the friction of the cases involving multi-dislocation motion. This complicated regime is out of our model and needs to be explored in the future.

## 4. Conclusion

In summary, this paper presented a semi-analytical model for the scale-dependent friction of a spherical tip sliding on a flat substrate. As contrast to the classical CM model, our MD simulation results demonstrate that the local friction coefficient in the slip zone is not a constant, but declines from a maximum at the stick/slip zone boundary till the contact edge. By introducing a dimensionless scale parameter $\beta$, a modified model is proposed to describe the contact shear stress in both the stick and slip zones. Based on this model, the global static friction coefficient is a constant when $\beta$ is smaller than 0.1, and decreases when $\beta$ increases further. For the asperity of small size with $\beta$ larger than 10, the static friction coefficient varies as the $-2/3$ power of $\beta$. By comparing with the numerical simulation results, it is found this model has successfully captured the unconventional features of shear stress distribution as well as the size dependency of static friction coefficient at nanoscale.

## Acknowledgement

Support from the National Natural Science Foundation of China (Grant No. 11525209) is acknowledged.

**Figure captions:**

Fig. 1 Schematic of the simulation model. $P$, $u_y$, $F$, $u_x$, and $2a$ indicate the normal load, the normal displacement, the lateral force, the lateral displacement of the tip, and the diameter of the contact area, respectively.

Fig. 2 The ratio of the lateral force to the normal, $F/P$, varies with the lateral displacement of the tip.

Fig. 3 (a) Distribution of local shear stress on contact surface, and (b) Variation of shear stress along five paths.

Fig. 4 Distribution of the local shear stress normalized by the local normal pressure along the path $\theta = 0°$ under different lateral displacement.

Fig. 5 Dependence of $1-\mu_l(a)/\mu_m$ on $(1-c^2/a^2)^{3/2}$ under different normal load.

Fig. 6 Dependence of $\beta$ on the dimensionless ratio $a/b$.

Fig. 7 Variation of the resultant lateral force with respect to the width of slip zone.

Fig. 8 Determination of the critical width of slip zone for different scale parameter $\beta$.

Fig. 9 Static friction coefficient varies with the dimensionless parameter $\beta$.

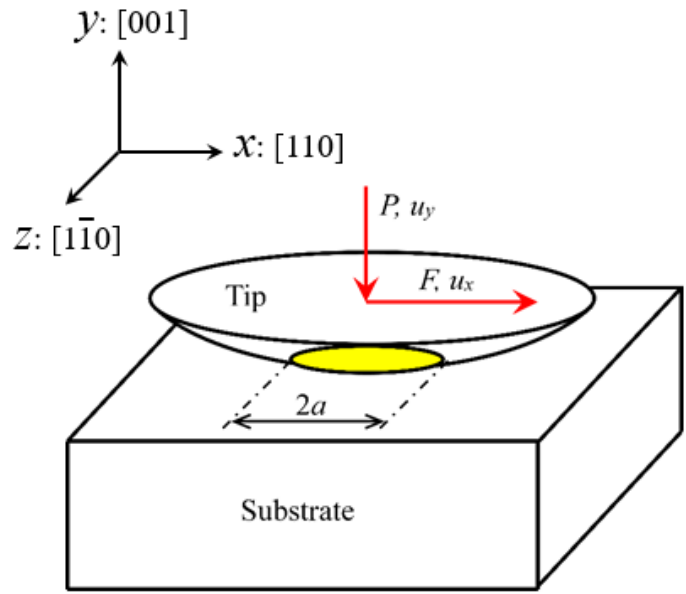

Fig. 1 Schematic of the simulation model. $P$, $u_y$, $F$, $u_x$, and $2a$ indicate the normal load, the normal displacement, the lateral force, the lateral displacement of the tip, and the diameter of the contact area, respectively.

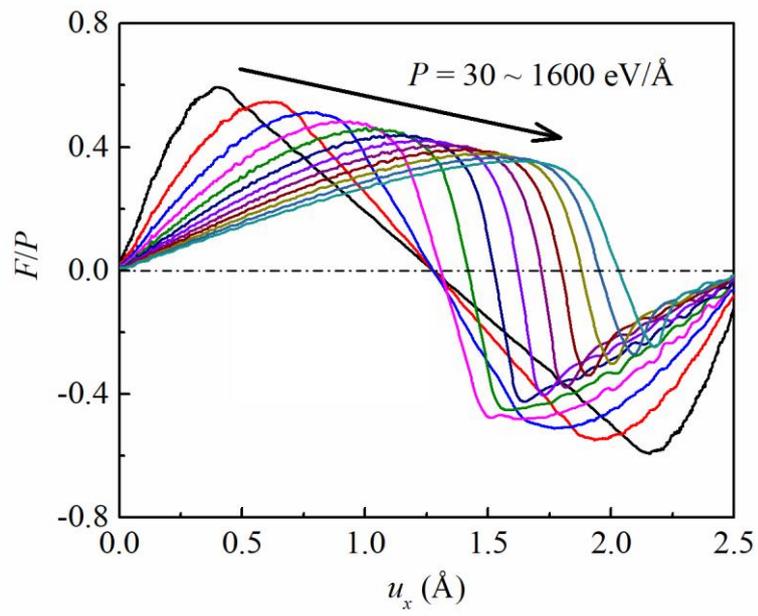

Fig. 2 The ratio of the lateral force to the normal, $F/P$, varies with the lateral displacement of the tip.

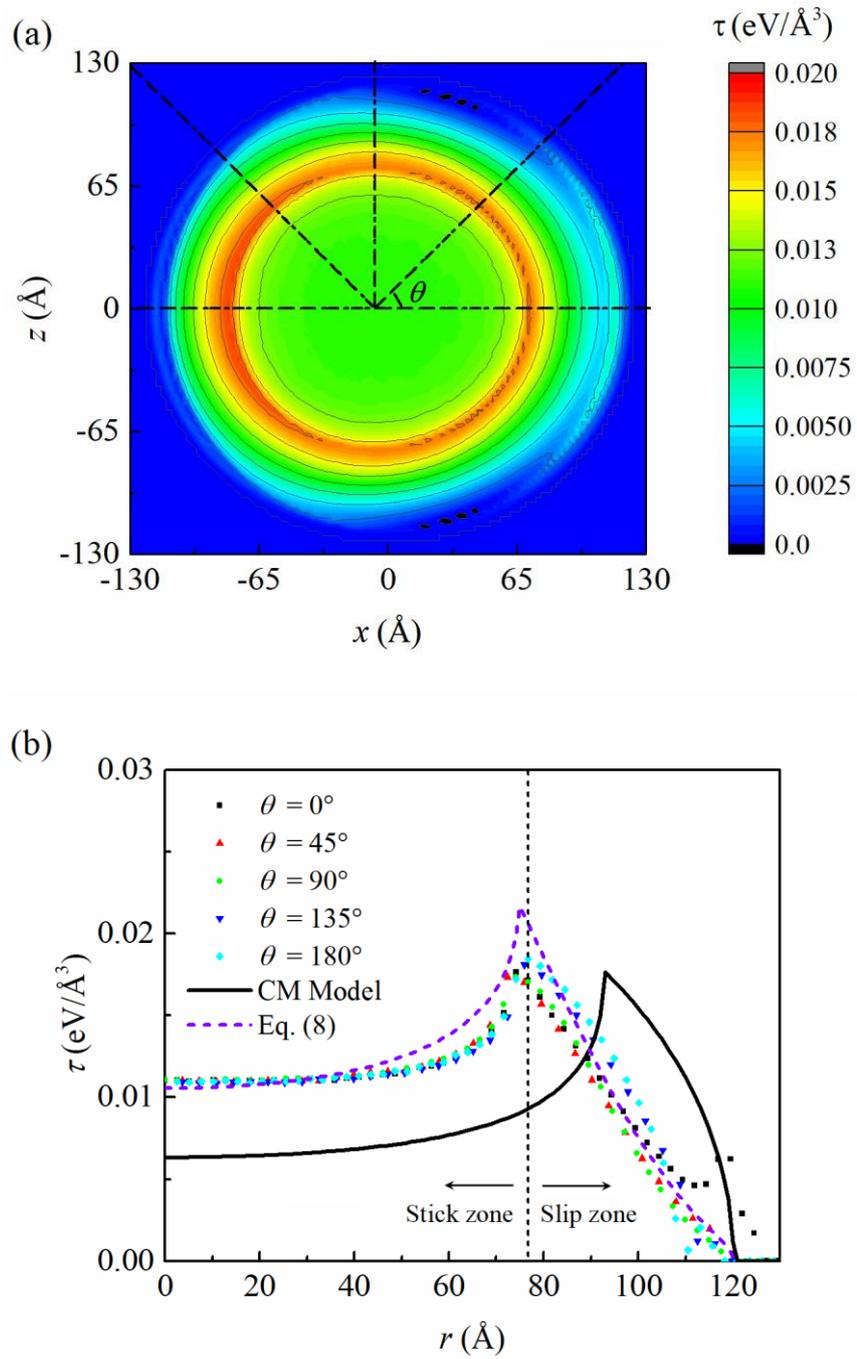

Fig. 3 (a) Distribution of local shear stress on contact surface, and (b) Variation of shear stress along five paths.

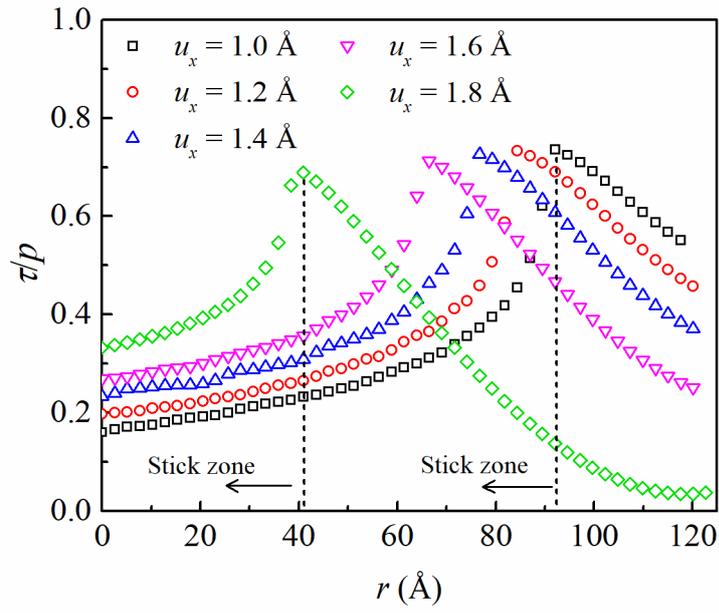

Fig. 4 Distribution of the local shear stress normalized by the local normal pressure along the path $\theta = 0°$ under different lateral displacement.

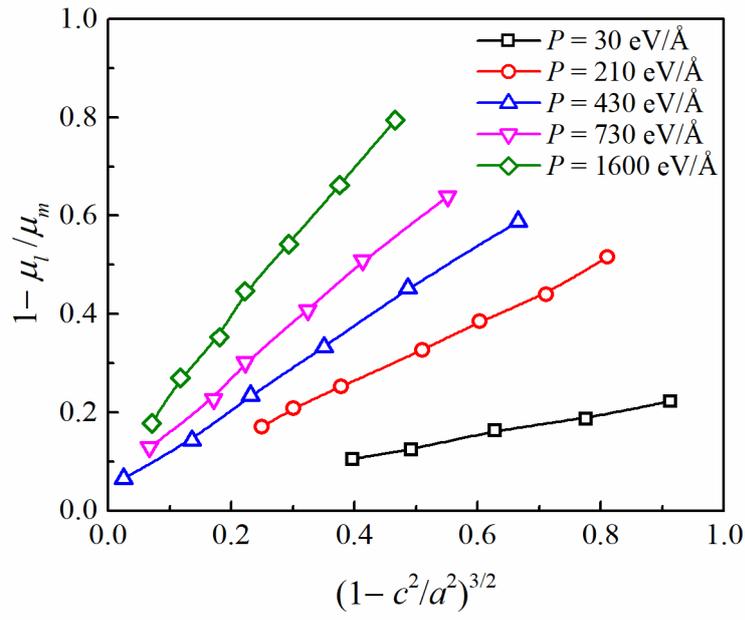

Fig. 5 Dependence of $1-\mu_l(a)/\mu_m$ on $(1-c^2/a^2)^{3/2}$ under different normal load.

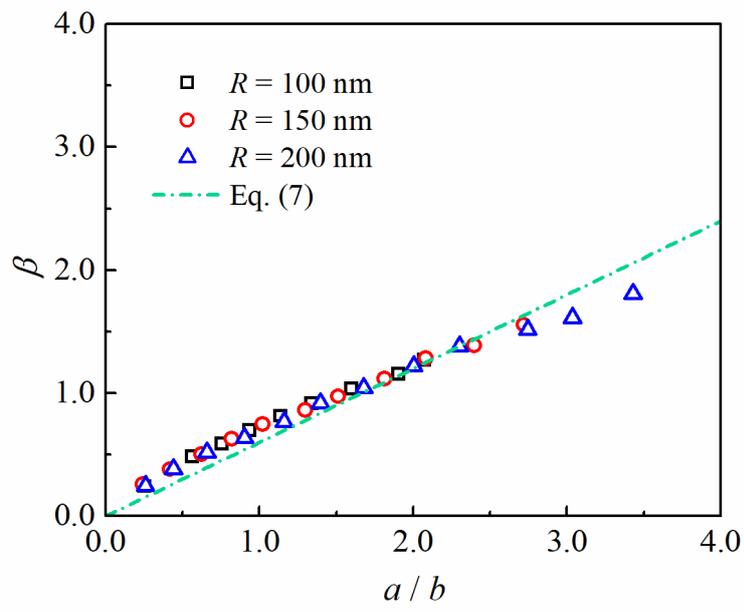

Fig. 6 Dependence of $\beta$ on the dimensionless ratio $a/b$.

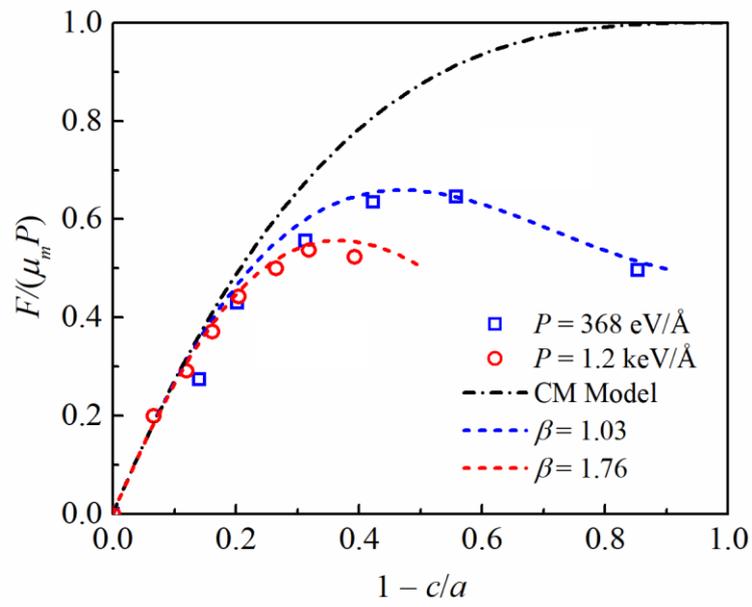

Fig. 7 Variation of the resultant lateral force with respect to the width of slip zone.

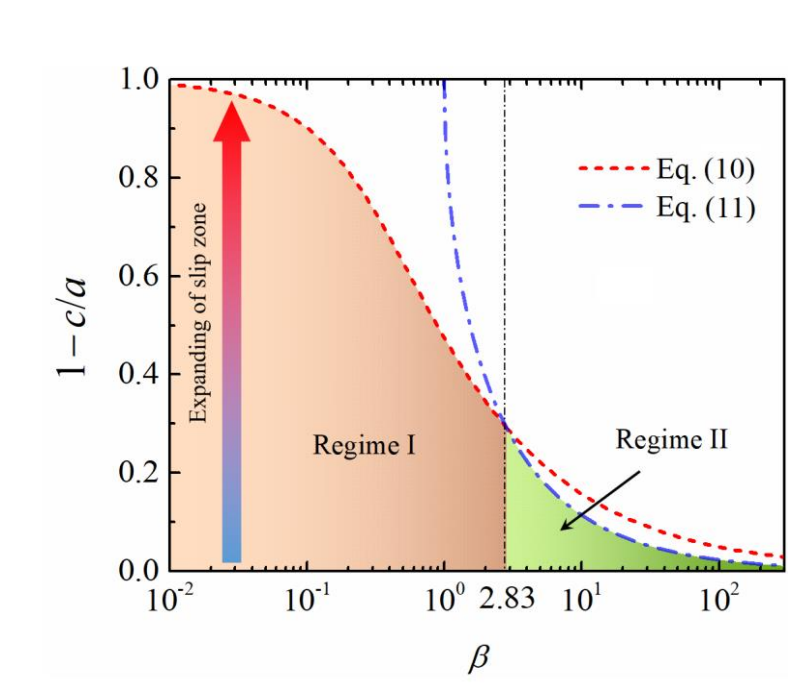

Fig. 8 Determination of the critical width of slip zone for different scale parameter $\beta$.

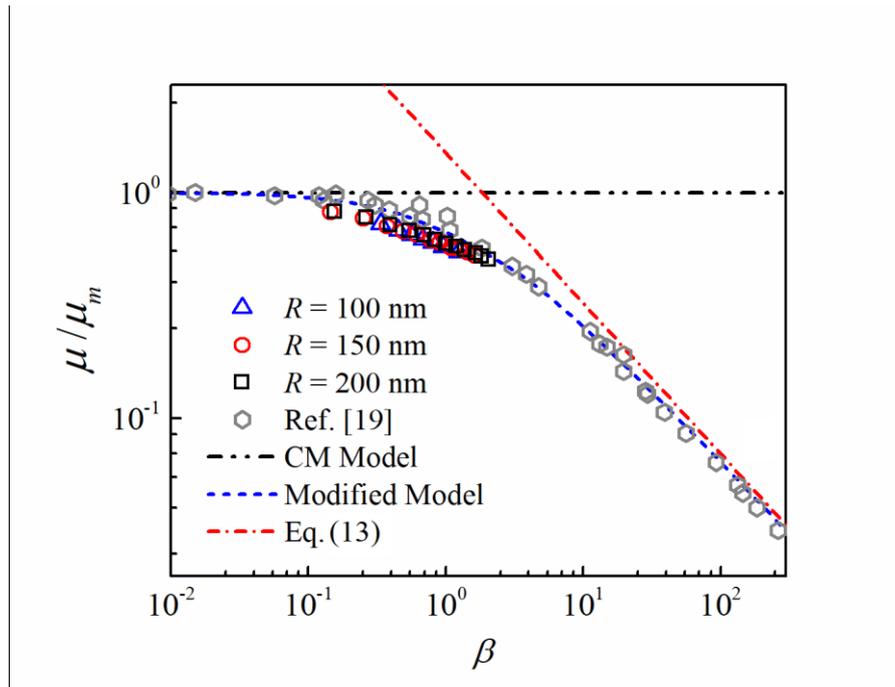

Fig. 9 Static friction coefficient varies with the dimensionless parameter $\beta$.